\documentclass[aps,twocolumn,showpacs,amsmath,amssymb,superscriptaddress,pre,longbibliography]{revtex4-2}

\usepackage{mathtools}
\usepackage{graphicx}
\usepackage{bm}
\usepackage{xcolor}
\usepackage{hyperref}
\hypersetup{hidelinks, colorlinks=true,linkcolor=blue,citecolor=blue}
\usepackage{comment}

\begin{document} 

\title{Fluctuation corrections to Lifshitz tails in disordered systems}

\author{Enrique Rozas Garcia}
\email{enrique.rozas.garcia@physics.gu.se}
\affiliation{Department of Physics, Gothenburg University, 41296 Gothenburg, Sweden}

\author{Johannes Hofmann}
\email{johannes.hofmann@physics.gu.se}
\affiliation{Department of Physics, Gothenburg University, 41296 Gothenburg, Sweden}
\affiliation{Nordita, Stockholm University and KTH Royal Institute of Technology, 10691 Stockholm, Sweden}

\date{\today}

\begin{abstract}
Quenched disorder in semiconductors induces localized electronic states at the band edge, which manifest as an exponential tail in the density of states. For large impurity densities, this tail takes a universal Lifshitz form that is characterized by short-ranged potential fluctuations. We provide both analytical expressions and numerical values for the Lifshitz tail of a parabolic conduction band including its exact fluctuation prefactor. Our analysis is based on a replica field integral approach, where the leading exponential scaling of the tail is determined by an instanton profile and fluctuations around the instanton determine the subleading preexponential factor. This factor contains the determinant of a fluctuation operator, and we avoid a full computation of its spectrum by using a Gel'fand-Yaglom formalism, which provides a concise general derivation of fluctuation corrections in disorder problems. We provide a revised result for the disorder band tail in two dimensions.
\end{abstract}

\maketitle

Disorder is an inherent property of physical systems, arising, for example, as defects or  impurities in doped semiconductors~\cite{VanMieghem1992}, frozen non-equilibrium degrees of freedom in quenched alloys and glasses~\cite{george2019high, mezard1987spin}, or thermal fluctuations~\cite{cusack1988physics}. It can also be engineered in photonic structures~\cite{Wiersma13,yu21} or quantum gases in an optical speckle potential~\cite{Roati08,Billy08}. 
For the electronic density of states in a semiconductor, disorder has two effects: First, it induces a narrowing of the band gap, where valence- and conduction-band levels are raised and lowered, respectively, by a characteristic impurity energy. 
Second, it causes band tailing, where fluctuations in the disorder potential give rise to localized states inside the band gap.
Describing the disorder-induced band narrowing and tailing is important because it affects semiconductor devices such as transistors~\cite{Shur84,VanMieghem1992,Khayer11} or limits the efficiency of solar cells~\cite{Jean17,wong2020impact}, and it sets the variable-range hopping conduction that dominates transport in localized systems at low temperatures~\cite{mott69}. 

In general, potential fluctuations that are large enough to generate states with energies far away from the band edge are rare, and the tail of localized levels takes an exponential form 
(written here for a single conduction band)~\cite{Economou1974, thouless1974electrons, mott1985minimum, VanMieghem1992},
\begin{equation}
\langle \rho(E) \rangle \overset{E \to -\infty}{=} A(E) e^{-B(E)} , \label{eq:dos_general}
\end{equation}
where $E$ is the energy below the band edge and~$\langle\cdot\rangle$ denotes a disorder average. 
Of particular interest is the limit of high impurity density, which is universal in the sense that the band tail is insensitive to details of the disorder correlations over a large energy range~\cite{VanMieghem1992}. This is the Lifshitz region or Lifshitz tail~\cite{lifshitz1963structure, lifshitz1964energy}, where~\eqref{eq:dos_general} takes a stretched exponential form [i.e., the argument $B(E)$ of the exponential has a power-law dependence on $E$]. Remarkably, even for the seemingly simple case of noninteracting particles in a Gaussian-correlated disorder potential, the disorder-averaged density of states can only be computed exactly in one space dimension~(1D)~\cite{halperin1965green}. In higher dimensions, dating back to works by Halperin and Lax~\cite{halperin1965green, Halperin1966, halperin1967impurity} and Zittartz and Langer~\cite{Zittartz1966}, the tail exponential $B(E)$ is obtained from a variational argument that determines the most likely disorder potential that gives rise to a bound state at large negative energy~$E$.
Such variational arguments are quite general~\cite{meibohm2023landau,meibohm2023caustics} and may be extended to include the effect of correlated disorder~\cite{john1984electronic}, magnetic fields~\cite{affleck1983two}, interactions~\cite{Yaida2016}, and also to describe the spectra of random operators~\cite{stollmann2001caught, khorunzhiy2006lifshitz, gebert2022lifshitz}. 
However, they do not capture the prefactor~$A(E)$ in the density of states~\eqref{eq:dos_general}, which is set by fluctuations of the disorder potential and which requires the evaluation of a functional determinant. The established result for~$A(E)$ derived by~\textcite{Brezin1980} is based on an analysis of large-order perturbations in~$\phi^4$ theory discussed in a classical paper by the same authors~\cite{Brezin1978}, which evaluates the fluctuation determinant by exact diagonalization~\cite{brezin77,Brezin1978}.
At the same time, the computation of fluctuation corrections is a recurring problem in other areas, for example to determine the decay rate of metastable states \cite{Marino2015}. Indeed, one of the first studies on the subject by~\textcite{gel1960integration,levit77} in the context of one-dimensional path integrals in quantum mechanics avoids the direct evaluation of the fluctuation spectrum by mapping the functional determinant to a differential equation, a more tractable problem compared to exact diagonalization. 

In this Letter, we show that the fundamental problem of computing the band tail prefactor~$A(E)$ of the Lifshitz tail~\eqref{eq:dos_general} has an efficient solution that does not rely on evaluating the full fluctuation determinant. This is achieved using a Gel'fand-Yaglom approach, which in particular provides a revised result for the band tail in two dimensions.
While the calculation is very general, we focus on the universal Lifshitz regime of infinitely dense but infinitesimally weak point scatterers, which is described by a Gaussian white noise disorder potential,
\begin{equation}
    \langle V(\bm{x}) \rangle = 0, \quad \langle V(\bm{x})V(\bm{x'}) \rangle = w^2
     \delta(\bm{x}-\bm{x'}) ,
    \label{eq:White_noise}
\end{equation}
and consider a parabolic conduction band described by the one-electron Hamiltonian
\begin{equation}
	H^V = - \frac{\hbar^2}{2m} \nabla^2 + V(\bm{x}) .
\label{eq:HV}
\end{equation}
Throughout this Letter, we set \mbox{$\hbar = 2m = 1$}.
We base our calculation on a 
functional integral representation for the disorder-averaged density of states, which we evaluate in a saddle-point approximation including fluctuations. The central results of our work are the following revised expressions for the density of states in the deep-tail regime~(i.e., for \mbox{$E\to - \infty$}),
\begin{equation}\label{eq:resultsbandtails}
\frac{\langle \rho(E) \rangle}{\rho_0(-E)} \overset{E \to -\infty}{=} \begin{cases} 8 \dfrac{|E|^{3/2}}{w^2} e^{-\frac{8 |E|^{3/2}}{3 w^2}}, & d=1, \\[3ex]
20.06~\dfrac{|E|^{3/2}}{w^{3}} e^{-11.70 \frac{|E|}{w^2}}, & d=2, \\[3ex]
8796\hspace{2.77779pt}~\dfrac{|E|}{w^{4}} e^{-37.79 \frac{|E|^{1/2}}{w^2}}, & d=3,
\end{cases}
\end{equation}
expressed here in relation to the density of states of the free Hamiltonian, 
\mbox{$\rho_0=E^{d/2-1}/(4\pi)^{d/2}\Gamma(d/2)$}. The 1D and 3D cases agree with previous results derived using other means~\cite{halperin1965green,Halperin1966,Brezin1980}, but the 2D result corrects an existing literature value~\cite{Brezin1980}.

The starting point of our analysis is a functional integral representation of the disorder-averaged density of states
\begin{equation}
\begin{split}
&\langle \rho(E) \rangle = \int \mathcal{D}[V] e^{- \int d\bm{x} \frac{V^2(\bm{x})}{2w^2}} \frac{1}{L^d}\sum_{n}\delta\left(E-E^V_n\right) \\
&= \frac{- i}{\pi L^d} \biggl\langle \frac{\partial}{\partial E} \bigl[\ln {\cal Z}^V(E + i0) - \ln {\cal Z}^V(E - i0)\bigr] \biggr\rangle. \label{eq:Greens_DOS}
\end{split}
\end{equation}
Here, the probability measure for the random field $V(\bm{x})$ is normalized to unity, $L^d$ is the $d$-dimensional volume, and $E^V_n$ is the $n$th energy eigenvalue for a realization of the Hamiltonian~\eqref{eq:HV} with potential $V(\bm{x})$. 
In the second line, we introduce the partition function
\begin{equation}
\begin{split}
{\cal Z}^V(E) &= \prod_n  \left(E_n^V-E\right)^{-1/2} \\
&= \int\mathcal{D}[\phi]\, \exp\left[-\frac{1}{2}\int d\bm{x} \, \phi\left(H^V - E\right)\phi\right] , \label{eq:ZV}
\end{split}
\end{equation}
which we express as a path integral over a scalar field~$\phi({\bf x})$. 
Written in this form, Eq.~\eqref{eq:Greens_DOS} is equivalent to the Edwards-Jones formula for the eigenvalue distribution of the random matrix $H^V$~\cite{edwards1976eigenvalue, livan2018introduction}. The factors $(E_n^V-E)^{-1/2}$ in the partition function have a branch cut for $E > E^V_n$, resulting in a discontinuity between complex energies $E\pm i0$, and thus a nonzero contribution to the density of states~\eqref{eq:Greens_DOS}.
The analytic continuation of the path integral to energies $E\pm{}i0$ is performed by rotating the fields $\phi$ in Eq.~\eqref{eq:ZV} into the complex plane by an angle $e^{\pm i\pi/4}$, i.e., on a contour $C^\pm$ along the positive and negative complex diagonals~\cite{callan1977fate,Cardy1978,ZinnJustin2002,Marino2015}. Note that with these contour choices, the integral representation~\eqref{eq:ZV} is convergent for all $E$ with nonzero imaginary part.

The disorder average~\eqref{eq:Greens_DOS} is difficult to compute directly due to the presence of the logarithm; therefore, we apply the replica trick~\cite{edwards1975theory}, 
\mbox{$\ln\mathcal{Z} = \lim_{N\to 0} (\mathcal{Z}^N - 1)/N$}, which avoids the logarithm in favor of introducing $N$ copies (replicas) of the original system. The disorder average is now performed exactly as a Gaussian integral over the potential $V(\bm{r})$, which gives
\begin{equation}
\langle \rho(E) \rangle = \lim_{N\to 0} \frac{(-E)^{1-\frac{d}{2}}}{i \pi w^2 N L^d}\int_{C^+ - C^-} \mathcal{D}[\Phi]\int d\bm{r} \, \Phi^2 e^{-\frac{1}{g} S[\Phi]} , \label{eq:evaluate}
\end{equation}
where the subscript indicates that we consider the difference over both contours, \mbox{$\Phi(\bm{r}) = \sqrt{-w^2/2E} \,  (\phi_1, \dots, \phi_N)$} is a vector of dimensionless replica fields that depend on a dimensionless position ${\bm r} = {\bm x} \sqrt{-E}$, and we introduce the scaling parameter 
\begin{equation}
\frac{1}{g} = \frac{(-E)^{2-\frac{d}{2}}}{w^2} . \label{eq:g}
\end{equation}
The effective action
\begin{equation}
S[\Phi] = \int d\bm{r}~\Phi \biggl( -\nabla^2 + 1 - \frac{1}{2}\Phi^2\biggr)\Phi
\label{eq:ADA_Action}
\end{equation}
corresponds to a $\phi^4$ Ginzburg-Landau theory \cite{Cardy1978}, where the disorder field is removed at the expense of introducing a nonlinear term. 
Crucially, for \mbox{$d<4$}, the prefactor $g^{-1}$ of the effective action~\eqref{eq:evaluate} is large in the deep-tail regime  \mbox{$E\to-\infty$}. 
This allows us to evaluate the integral over the replica field~$\Phi$ in the saddle-point approximation~\cite{Bender1999}, that is, by expanding around the dominant contribution to the action. 

\begin{figure}[t]
\includegraphics[width=0.8\linewidth]{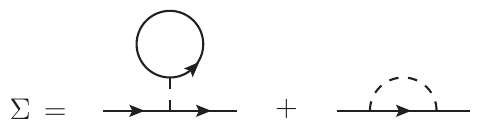}
\caption{Divergent self-energy diagrams for the $\phi^4$ theory~\eqref{eq:ADA_Action}.}
\label{fig:1}
\end{figure}

Before evaluating the saddle point, however, note that the theory~\eqref{eq:ADA_Action} requires renormalization. 
The divergent one-loop self-energy corrections are shown in Fig.~\ref{fig:1}, where a continuous line denotes the free Green's function and the vertex is split to connect two replica indices. Physically, the divergences are a consequence of the white-noise approximation~\eqref{eq:White_noise} to the disorder potential, which breaks down as the field $\Phi$ varies over length scales smaller than the separation between scatterers. The divergences are subtracted by adding a counterterm to the action~\eqref{eq:ADA_Action},
\begin{equation}
\delta S[\Phi] = \frac{N+2}{2} g \, G_0 \int d\bm{r} \, \Phi^2 ,
\label{eq:deltaS}
\end{equation}
which is linear in $g$ and will thus give a subleading correction in the saddle-point approximation. Here,
\begin{equation}
G_0 = \int \frac{d\bm{q}}{(2\pi)^d} \, \frac{1}{q^2 + 1} \label{eq:G0}
\end{equation}
is the free Green's function at coinciding points, which diverges in $d\geq 2$, and the factor $N+2$ in Eq.~\eqref{eq:deltaS} arises from the different contractions of the replica fields. 
The self-energy $\Sigma(E)$ computed including the counterterm is finite and represents the band narrowing that shifts the reference point $E_0$ of the energy scale. Including the finite part of $G_0$ in the counterterm corresponds to a renormalization choice $\Sigma(E)|_0=0$ that puts this reference at the origin~\cite{Altland2010}. Another possible choice is $\partial\Sigma(E)/\partial E |_0 = 1$, which is motivated from a coherent potential approximation~\cite{Brezin1980, Thouless1978}. Results computed in different schemes are identical and linked by a shift in the energy parameter (with possible logarithmic corrections in 2D).

\begin{figure}[t]
\centering
\hspace{-20pt}\includegraphics{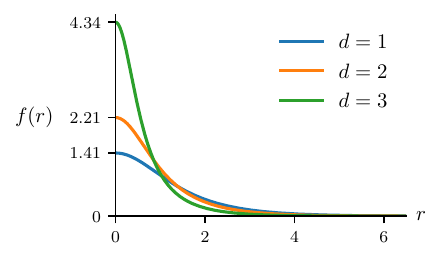}
\caption{Instanton solution of the saddle-point equation~\eqref{eq:Instanton}, which determines the leading scaling of the band tail~$B(E)$.}
\label{fig:2}
\end{figure}

The leading saddle point of the action~\eqref{eq:ADA_Action} is the trivial configuration~\mbox{$\Phi = 0$}; however, this contribution is the same for both contours $C^{\pm}$ and cancels when taking the difference in Eq.~\eqref{eq:evaluate}. 
The parameter $B(E)$ in the exponential band tail~\eqref{eq:dos_general} is thus determined by the action at the subleading saddle point of Eq.~\eqref{eq:evaluate}, which is of order \mbox{$S[\Phi] = O(g^{-1})$}.
The field configuration that minimizes the effective action~\eqref{eq:ADA_Action} is of the form \mbox{$\Phi(\bm{r} + \bm{r}_0) = f(r) \, \bm{u}$}~\cite{Coleman1978, ZinnJustin2002}, where $\bm{r}_0$ is an arbitrary center, $\bm{u}$ an arbitrary unit vector in replica space, and $f(r)$ is a nodeless radial function known as the instanton. The saddle-point equation is
\begin{equation}
-\nabla^2 f(r) + f(r) - f^3(r) = 0, \label{eq:Instanton}
\end{equation}
with boundary conditions \mbox{$f'(0) = f(r\to\infty) = 0$}. 
We can interpret it as a Schrödinger equation for the potential $-f^2(r)$ that describes the most likely shape of the disorder potential with a bound state at energy $E$~\cite{Yaida2016}. 
Figure~\ref{fig:2} shows the instanton profile in $d=1,2$, and $3$. In 1D, the solution \mbox{$f(r) = \sqrt{2}\, \text{sech}(r)$} is known analytically, and for $d\geq 2$, we employ a spectral renormalization method to solve Eq.~\eqref{eq:Instanton} numerically~\cite{suppl}. 
The exponential band tail is then given by the saddle-point action
\begin{equation}
B(E) = \frac{1}{g} S[f] = \frac{I_4}{2 g} 
, \label{instanton_action}
\end{equation}
where
\begin{equation}
I_n = \int d\bm{r} \, f^n(r) . \label{eq:moment}
\end{equation}
The corresponding values for $I_4$ are well established~\cite{brezin77,ZinnJustin2002} and tabulated in Table~\ref{tab:1}.

\begin{table}[t]
\begin{ruledtabular}
\begin{tabular}{llll} 
$d$ & $1$ & $2$ & $3$ \\
\hline
$I_2$ & 4 & \hspace{-4.62497pt}11.7009 & \hspace{-4.62497pt}18.8973 \\
$I_4$ & \hspace{-4.62497pt}16/3 & \hspace{-4.62497pt}23.4018 & \hspace{-4.62497pt}75.5890 \\
$e^{I_2G_0} D\left(1/3\right)$  & 1/4 & 1.0558 & 0.364 \\
$e^{3I_2G_0} D(1)$  & \hspace{-7.7778pt}$-$1/12 & \hspace{-7.7778pt}$-$5.35 & \hspace{-7.7778pt}$-$0.0096 \\
\end{tabular}
\end{ruledtabular}
\caption{Numerical values for the moments of the instanton profile~\eqref{eq:moment} and for the normalized longitudinal and transverse fluctuation determinant $e^{3zI_2G_0} D(z)$ in \mbox{$d=1$, $2$},~and~$3$. 
}
\label{tab:1}
\end{table}
    
Our main task in this Letter is to determine the leading contribution to the prefactor $A(E)$ of the exponential band tail~\eqref{eq:dos_general} by evaluating the next order in the saddle-point approximation. To this end, we expand the action to second order in new fields that represent longitudinal ($\delta f_\parallel$) and transverse ($\delta f_{\perp}^k$) fluctuations,
\begin{equation}
\Phi(\bm{r}) = \bm{u} \left[f(r) + \delta f_\parallel(\bm{r})\right] + \sum_{k = 1}^{N-1} \bm{v}_{k} \delta f_{\perp}^{k}(\bm{r}), 
\label{eq:shift}
\end{equation}
where the set $\{\bm{u}, \bm{v}_1,\dots, \bm{v}_{N-1}\}$ forms an orthonormal basis of replica space. 
However, certain fluctuations do not represent small corrections to the saddle-point action.
These so-called zero modes correspond to changes of the orientation~$\bm{u}$ of the instanton in replica space and of its center~$\bm{r}_0$. 
These changes leave the action invariant and thus cannot be treated perturbatively but must be integrated exactly. This is done using the method of collective coordinates~\cite{Coleman1985, rajaraman1982solitons}, resulting in an overall factor
\begin{equation}
A_0 = \left(\frac{I_4}{4\pi  g}\right)^{\frac{d}{2}} \left[L \sqrt{-E}\right]^d \,
\left(\frac{I_2}{\pi  g}\right)^{\frac{N-1}{2}} \frac{2\pi^{N/2}}{\Gamma(N/2)} .
\label{eq:A0}
\end{equation}
Here, terms in parenthesis are the Jacobian of the transformation to the collective position~$\bm{r}_0$ and angle coordinate~$\bm{u}$. Note that this contribution introduces additional factors of $g$ and thus contributes to the energy dependence of the prefactor $A(E)$. The remaining terms correspond to the measure of the degenerate symmetry spaces, which is the $N$-dimensional surface element in replica space and the real-space volume $L^d$. 

Formally, the full fluctuation correction reads
\begin{equation}
\begin{split}
&A(E) = \lim_{N\to 0} \frac{(-E)^{1-\frac{d}{2}}}{i \pi w^2 N L^d} I_2 A_0 \, e^{-\frac{N+2}{2} G_0 I_2} \\[1.5ex]
& \times \int \mathcal{D}[\delta f] \, e^{- \frac{1}{g} \int d\bm{r} \left(\delta f_\parallel \Delta_1 \delta f_\parallel + \sum_{k} \delta f_{\perp}^{k} \Delta_{1/3} \delta f_{\perp}^{k}\right)}  ,
\end{split} 
\label{eq:fluctuations}
\end{equation}
where the factor $I_2$ is the saddle-point value of the field $\Phi^2$ in Eq.~\eqref{eq:evaluate}, the exponential prefactor is the counterterm contribution~\eqref{eq:deltaS}, and we define the fluctuation kernel
\begin{equation}
\Delta_z = -\nabla^2 + 1 - 3zf^2 , \label{eq:deltaz}
\end{equation}
which describes the decoupled longitudinal (\mbox{$z=1$}) and transverse (\mbox{$z=1/3$}) fluctuations. 
The excluded zero modes that correspond to an infinitesimal shift along one coordinate axis \mbox{$\alpha = 1,\ldots,d$} are linked to~$d$ zero modes of the longitudinal fluctuation kernel~$\Delta_1$ given by \mbox{$\partial_\alpha f(\bm{r})$}. 
Likewise, the operator $\Delta_{1/3}$ 
has a zero mode \mbox{$f(\bm{r})$} corresponding to rotations in replica space. 
The path integral in Eq.~\eqref{eq:fluctuations} over the fluctuations $\delta f_\parallel$ and $\delta f_{\perp,k}$ excludes the zero modes. It is then a Gaussian integral and results in factors involving the functional determinant of~$\Delta_z$,
\begin{equation}
D(z) = \lim_{\varepsilon\to 0}\frac{1}{\varepsilon^m}\det \frac{\Delta_z + \varepsilon}{\Delta_0 + \varepsilon} 
, \label{eq:D_z}
\end{equation}
where for the shifted operator $\Delta_z+\varepsilon$ the zero modes have eigenvalue $\varepsilon$, and are thus removed in the limit \mbox{$\varepsilon\to 0$} ($m$ is the degeneracy of zero modes, with $m=d$ for $z=1$ and $m=1$ for $z=1/3$).

Taken together, the longitudinal and transverse fluctuation contribution to the path integral~\eqref{eq:fluctuations} reads
\begin{equation}
\begin{split}
A(E) &= 
\lim_{N\to 0} \frac{(-E)^{1-\frac{d}{2}}}{\pi w^2 N L^d} I_2 A_0\\
&\qquad \times  \Bigl[e^{3 I_2 G_0} \bigl|D(1)\bigr|\Bigr]^{-\frac{1}{2}}
\Bigl[e^{I_2 G_0} D\left(\tfrac{1}{3}\right)\Bigr]^{-\frac{N-1}{2}} ,
\end{split} \label{eq:Afluc}
\end{equation}
where the exponent in the third factor accounts for the \mbox{$N-1$} transverse replica field directions, and we split up the counterterm contribution. 
Note that the longitudinal kernel $D(1)$ is negative since it contains a single eigenfunction $f$ with negative eigenvalue \mbox{$-2$}. The square root is still well defined after analytic continuation, and the resulting imaginary unit cancels with that in the prefactor of Eq.~\eqref{eq:fluctuations}. 
Moreover, in \mbox{$d>1$}, the fluctuation determinant~$D(z)$ diverges. 
The mathematical origin of this divergence is identified by expanding the determinant in powers of $z$,
\begin{equation}
\text{det}\biggl[1 - \frac{3zf^2}{-\nabla^2 + 1}\biggr]  = e^{1 - 3 z I_2 G_0 + O(z^2)} ,
\label{eq:z_expansion}
\end{equation}
where $G_0$ is introduced in Eq.~\eqref{eq:G0} and higher-order terms in $z$ are finite. 
The divergent term in Eq.~\eqref{eq:z_expansion}, however, is precisely canceled by the leading-order counterterm~\eqref{eq:deltaS}, rendering the fluctuation determinant~\eqref{eq:Afluc} manifestly finite.

To evaluate the finite renormalized fluctuation determinant, one could in principle compute the eigenvalues of $\Delta_z$  and evaluate their product, which was done in a similar context by~\textcite{Brezin1978} to determine the large-order behavior of perturbation series in $\phi^4$ theories. Here, we make use of a more direct method using spectral functions. This method was originally proposed by Gel'fand and Yaglom~\cite{gel1960integration}, and applied to higher-dimensional problems with radial symmetry in Ref.~\cite{Dunne2006} (for a review, see Refs.~\cite{Kirsten2004, Dunne2008, Kirsten2008}). Crucially, the Gel'fand-Yaglom method does not require knowledge of the eigenvalues, but instead expresses the functional determinant in terms of the solution to an ordinary differential equation, which greatly simplifies the overall calculation. 

\begin{figure}[t!]
\centering
\includegraphics[width=1\linewidth]{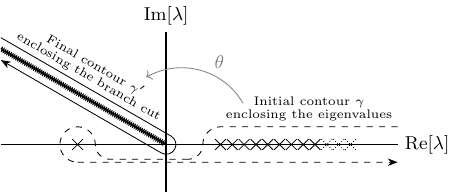}
\caption{Complex structure of the integral~\eqref{eq:Contour_1}. Crosses indicate eigenvalues (shown here for longitudinal fluctuations with one negative eigenvalue) and the wavy line denotes the branch cut. The dashed line indicates the initial integration contour $\gamma$ that encloses all eigenvalues, and the solid line shows the deformed contour $\gamma'$ that runs around the  branch cut.}
\label{fig:3}
\end{figure}

The Gel'fand-Yaglom method is based on the formal identity~\cite{levit77}
\begin{equation}
\det \Delta_z = \prod_n \lambda_{n} = e^{-\zeta_z'(0)}, \label{eq:GY_formal}
\end{equation}
where $\lambda_{n}$ are the eigenvalues 
of the operator $\Delta_z$, defined in Eq.~\eqref{eq:deltaz}, and 
\begin{equation}
\zeta_z(s) = \sum_n \frac{1}{\lambda_{n}^s}. \label{eq:zeta_function}
\end{equation}
We introduce an analytic function $\mathcal{F}_z(\lambda)$ with zeros that coincide in location and multiplicity with the eigenvalues~$\lambda_n$, in terms of which
\begin{equation}
\begin{split}
&\zeta_z(s) - \zeta_0(s) = \frac{1}{2\pi i}\int_\gamma d\lambda\, \lambda^{-s} \frac{d}{d\lambda}\left[\ln\frac{\mathcal{F}_z(\lambda)}{\mathcal{F}_0(\lambda)}\right] \\ &=e^{-is(\pi-\theta)}\frac{\sin \pi s}{\pi} \int_0^\infty d\lambda \, \lambda^{-s} \frac{d}{d\lambda} \left[\ln\frac{\mathcal{F}_z(e^{i\theta} \lambda)}{\mathcal{F}_0(e^{i\theta} \lambda)}\right]. \label{eq:Contour_1}
\end{split}
\end{equation}
Here, the contour $\gamma$ encircles all eigenvalues, and we place the branch cut of the integrand at an angle~$\theta$ with the real axis to avoid overlapping with the eigenvalues~\cite{Kirsten2004}; see Fig.~\ref{fig:3} for an illustration. 
To obtain the last identity, we deform the contour to enclose the branch cut, which is shown as a continuous line in Fig.~\ref{fig:3}. We may now take the derivative with respect to $s$ to obtain
\begin{equation}
    \det \frac{\Delta_z}{\Delta_0} 
    = e^{-\zeta_z'(0) + \zeta_0'(0)} 
    =\frac{\mathcal{F}_z(0)}{\mathcal{F}_0(0)} =  \prod_{l=0}^\infty \left[\frac{\mathcal{F}^{(l)}_z(0)}{\mathcal{F}^{(l)}_0(0)}\right]^{{\rm d}(l; d)},
    \label{eq:Contour_2}
\end{equation}
where we use the rotational symmetry of~$\Delta_z$ to factorize the determinant and by extension~$\mathcal{F}$ into angular components $\mathcal{F}^{(l)}$~\cite{Dunne2006}, and we denote the angular degeneracy of the eigenvalues by ${\rm d}(l; d)$. 
Note that in 1D the rotational symmetry reduces to a parity symmetry; thus, $l$ labels only nondegenerate even ($l=0$) and odd ($l=1$) contributions.
Intuitively, the function $\mathcal{F}_z(\lambda)$ generalizes the characteristic polynomial of finite-dimensional matrices to the operator $\Delta_z$. Equation~\eqref{eq:Contour_2} then corresponds to the result that the constant term in the characteristic polynomial is the determinant.

To obtain $\mathcal{F}^{(l)}_z(\lambda)$, we follow the same heuristic as used to find eigenvalues numerically with the shooting method~\cite{Kirsten2004}: Consider the function \mbox{$\phi^{(l)}_z(\lambda, r)$} obtained by solving the initial value problem
\begin{equation}
\left(\Delta^{(l)}_z + \varepsilon\right)\phi_z^{(l)} = \lambda \phi_z^{(l)}, \quad \phi_z^{(l)}(\lambda, r) \overset{r\to 0}{\sim} r^{l+\frac{d-1}{2}}, \label{eq:bvp}
\end{equation}
where we include the factor $\varepsilon\ll 1$ in Eq.~\eqref{eq:D_z}, and 
\begin{equation}
\Delta^{(l)}_z = -\frac{d^2}{dr^2} + \frac{(l+\frac{d-3}{2}) (l + \frac{d-1}{2})}{r^2} + 1 -3z f^2.
\label{eq:Radial_operator}
\end{equation}
If $\lambda$ is such that the function vanishes at infinity, \mbox{$\phi_z^{(l)}(\lambda, \infty)=0$}, $\phi$ will also be an eigenfunction of~$\Delta^{(l)}_z$ with eigenvalue $\lambda$. But this is exactly the defining property of $\mathcal{F}_z^{(l)}$, and hence we set \mbox{$\mathcal{F}^{(l)}_z (\lambda) = \phi_z^{(l)}(\lambda, \infty)$}. 
Note that only the ratio $\mathcal{R}_{z,\varepsilon}^{(l)} = \mathcal{F}_z^{(l)} /\mathcal{F}_0^{(l)}$ contributes to Eq.~\eqref{eq:Contour_2}, so it is useful to derive an equation for $\mathcal{R}_{z,\varepsilon}^{(l)}$ directly~\cite{Dunne2005beyond},
\begin{equation}
\begin{split}
&0 = \mathcal{R}_{z,\varepsilon}^{(l)\prime\prime}(r) +  3zf^2(r) \mathcal{R}_{z,\varepsilon}^{(l)} (r)  \\
&+ \left[ \frac{2l + d  - 1}{r} + 2 \sqrt{1+\varepsilon} \frac{I_{l+\frac{d}{2}}\left(\sqrt{1+\varepsilon} r\right)}{I_{l+\frac{d}{2}-1}\left(\sqrt{1+\varepsilon} r\right)} \right] \mathcal{R}_{z,\varepsilon}^{(l)\prime}(r) ,\label{eq:R_ODE}
\end{split}
\end{equation}
with boundary conditions \mbox{$\mathcal{R}_{z,\varepsilon}^{(l)}(0)=1$} and \mbox{$\mathcal{R}_{z,\varepsilon}^{(l)\prime}(0) = 0$}, inherited from Eq.~\eqref{eq:bvp}. Here, $I$ is the modified Bessel function of the first kind. 

The renormalized determinant, finite to all orders in $z$ and without zero modes, reads
\begin{equation}
\begin{split}
& \ln \bigl[e^{-z I_2 G_0} D(z)\bigr]
\\
& = \lim_{\varepsilon\to 0} \sum_{l=0}^\infty {\rm d}(l; d) \left[\frac{1}{\varepsilon^{\delta_l}}\ln \mathcal{R}^{(l)}_{z,\varepsilon}(\infty) - z r^{(l)}\right], \label{eq:det_final}
\end{split}
\end{equation}
where $\delta_l$ is equal to unity if $\Delta_z^{(l)}$ has a zero mode, and zero otherwise. The second term in the brackets is the counterterm contribution, which in our discussion of Eq.~\eqref{eq:z_expansion} we identified as the term of linear order in $z$, and which is thus written as
\begin{equation}
\begin{split}
r^{(l)} &= \frac{\partial \ln \mathcal{R}^{(l)}_{z,\varepsilon}(\infty)}{\partial z} \biggr\rvert_{z=0} \\
&= -3\int_0^\infty dr~r f^2(r) K_\nu(r)I_\nu(r) ,
\end{split} \label{eq:counterterm_def}
\end{equation}
where \mbox{$\nu=l+d/2-1$}. In the last identity we used the connection between $\mathcal{R}$ and the established asymptotic expansion of the Jost function in scattering theory~\cite{Dunne2006}. 
The values of the renormalized functional determinant~$e^{3zI_2G_0} D(z)$, Eq.~\eqref{eq:det_final}, are listed in Table~\ref{tab:1}. In the 1D case, the results agree with the known exact values derived from the eigenvalues of the Pöschl-Teller potential~\cite{Marino2015}. The calculations for two and three dimensions were previously carried out by~\textcite{Brezin1978} by a direct computation of the spectrum of $\Delta_z$. We agree with their results in 3D, but not in the 2D case. 
However, repeating their calculation in 2D, which also illustrates the comparative efficiency of the Gel'fand-Yaglom approach, we confirm the result derived in this Letter~\cite{suppl}. 

Taking everything together, our final result for the density of states is (taking the replica limit $N\to0$)
\begin{equation}
\begin{split}
\frac{\langle \rho(E) \rangle}{\rho_0(-E)}   &\overset{E \to -\infty}{=} \Gamma(d/2)
\left(\frac{I_2 I_4^d}{\pi} \frac{D(1/3)}{|D(1)|} \right)^\frac{1}{2}
g^{-\frac{d+1}{2}}
e^{-\frac{1}{2g}I_4} .
\end{split}
\end{equation}
The numerical results for these band tails are stated in Eq.~\eqref{eq:resultsbandtails}. 
Again, our results agree with the explicit result in 1D~\cite{VanMieghem1992}. In 2D, we correct the values for the prefactor reported in Refs.~\cite{Brezin1978, Brezin1980}, and in 3D we agree with Ref.~\cite{Brezin1980} when adjusting for a factor $e^{-I_4/16\pi}$ that accounts for a different renormalization choice (i.e., a shifted band origin).

In summary, we have quantified disorder effects on the electronic density of states in the universal Lifshitz regime including fluctuations, which set the prefactor of the Lifshitz tail.
Using a replica path-integral approach, we derive the leading exponential form of the tail from an
instanton saddle-point profile, which describes the most likely disorder potential that gives rise to a particular deep bound state. Importantly, we also obtain the fluctuation correction to this result, which sets the magnitude of this tail. Here, the main technical advance of our work is a generalization of the Gel'fand-Yaglom approach to evaluate the fluctuation determinant around the saddle-point solution, which avoids a direct evaluation of the fluctuation spectrum. We thus provide an and efficient calculation of disorder tails that confirms results in 1D and 3D obtained using other means, and which corrects a result for the important case of two-dimensional systems. The derivation presented in this Letter has the dual advantage of simplicity and adaptability to more complex scenarios. For instance, extending our calculations to describe a broader class of band structures with, for example, fourth-order corrections, spin-orbit coupling, or band structure asymmetries, or to include more general disorder correlations are avenues for future work.

\begin{acknowledgments}
We thank Jan Meibohm for discussions. 
This work is supported by Vetenskapsr\aa det (Grant No. 2020-04239).
\end{acknowledgments}

\bibliography{bib_disorder}

\end{document}


\title{Supplemental Material for ``Fluctuation corrections to Lifshitz tails in disordered systems"}

\author{Enrique Rozas Garcia}
\email{enrique.rozas.garcia@physics.gu.se}
\affiliation{Department of Physics, Gothenburg University, 41296 Gothenburg, Sweden}

\author{Johannes Hofmann}
\email{johannes.hofmann@physics.gu.se}
\affiliation{Department of Physics, Gothenburg University, 41296 Gothenburg, Sweden}
\affiliation{Nordita, Stockholm University and KTH Royal Institute of Technology, 10691 Stockholm, Sweden}

\date{\today}
\maketitle

This supplemental material expands on the discussion in the main text and presents a numerical method used to compute the instanton profile (Sec.~\ref{sec:SM_spectral_renormalization}), an alternative computation of the fluctuation determinant based on an exact diagonalization and extrapolation of its spectrum (Sec.~\ref{app:spectrum}), and useful numerical details in the implementation of the Gel'fand-Yaglom approach (Sec.~\ref{appendices:R-nitty-gritty}).

\section{Spectral renormalization}
\label{sec:SM_spectral_renormalization}

The saddle-point equation for the instanton configuration, Eq.~\eqref{main-eq:Instanton} of the main text, is commonly solved using the shooting method. Here, we discuss the spectral renormalization method proposed by~\textcite{Ablowitz2005} as a rapidly converging and robust alternative. It is based on the mapping between functions implied by the Fourier transform of~\eqref{main-eq:Instanton}
\begin{equation}
\tilde{f}_{n+1}(\bm{k}) = \frac{\tilde{f_n^3}(\bm{k})}{1 + \bm{k}^2}. \label{eq:bad_map}
\end{equation}
Fixed points of this map are solutions of Eq.~\eqref{main-eq:Instanton}. Unfortunately, a direct iteration of this equation does not usually converge to the instanton profile $f(r)$, but to different and undesirable fixed points (e.g., zero or infinity). 
To improve the convergence, we follow~\cite{Ablowitz2005} and impose an additional integral condition characteristic of the fixed points of~\eqref{eq:bad_map},
\begin{equation}
\int d\bm{k}\, \bigl|\tilde{f}(\bm{k})\bigr|^2  = \int d\bm{k} \, \tilde{f}^*(\bm{k}) \frac{\tilde{f^3}(\bm{k})}{1 + \bm{k}^2}. \label{eq:Renormalization_restriction}
\end{equation}
This is done by rescaling the function \mbox{$f_n\to \gamma_n f_n$} at each iteration, where the value of $\gamma_n$ is chosen to fulfill ~\eqref{eq:Renormalization_restriction}
\begin{equation}
\gamma_n^2 = \frac{\int d\bm{k}\, \bigl|\tilde{f}(\bm{k})\bigr|^2}{\int d\bm{k} \, \tilde{f}^*(\bm{k}) \frac{\tilde{f^3}(\bm{k})}{1 + \bm{k}^2}}.
\end{equation}
The rescaled map is given by
\begin{equation}
\tilde{f}_{n+1}(\bm{k}) = \gamma_n^{\frac{3}{2}}\frac{\tilde{f_n^3}(\bm{k})}{1 + \bm{k}^2}. \label{eq:spectral_renormalization}
\end{equation}
For general instanton profiles, the fixed-point map~\eqref{eq:spectral_renormalization} is evaluated using a fast Fourier transform, which converges to the desired solution within a small number of iterations. 
An additional simplification in our case comes from the observation that $f(r)$ has rotational symmetry. This property is preserved by both the map~\eqref{eq:spectral_renormalization} and the Fourier transform. For rotationally symmetric functions, the Fourier transform simplifies to the one-dimensional integral
\begin{equation}
    \tilde{f}(k) = \frac{(2 \pi)^{d/2}}{k^{d/2-1}}
\int_{0}^{\infty}dr \, r^{d/2}f(r)J_{d/2-1}(kr), \label{eq:rotational_FT}
\end{equation}
where $J$ is the Bessel function of the first kind. This integral is numerically evaluated using a Gauss-Legendre quadrature rule. The results shown in Fig.~\ref{main-fig:1} and Tab.~\ref{main-tab:1} of the main text are obtained using the spectral renormalization iteration~\eqref{eq:spectral_renormalization} with an initial Gaussian profile. 

\section{Direct diagonalization of the fluctuation determinant}\label{app:spectrum}

The results for the fluctuation determinant that we derive in the main paper using the Gel'fand-Yaglom method agree with the exact result in \mbox{$d=1$} and also with the result in \mbox{$d=3$} obtained by~\textcite{Brezin1978}. However, Ref.~\cite{Brezin1978} also states results in \mbox{$d=2$}, and our result differs from this work. To resolve this discrepancy and to confirm our result, this supplemental section presents an alternative derivation in $d=2$ following~\cite{Brezin1978}, which is based on an evaluation of the complete spectrum of the kernels of longitudinal and transverse fluctuations. The results obtained in this section agree with our calculation in the main text and correct the values reported in~\cite{Brezin1978}. Furthermore, it highlights the use of the Gel'fand-Yaglom approach, which avoids the explicit computation of the fluctuation spectrum.

\begin{figure}[t!]
\scalebox{0.65}{\includegraphics{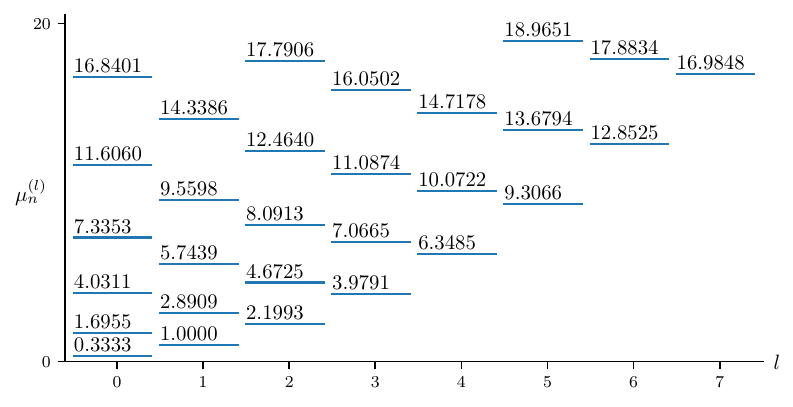}\includegraphics{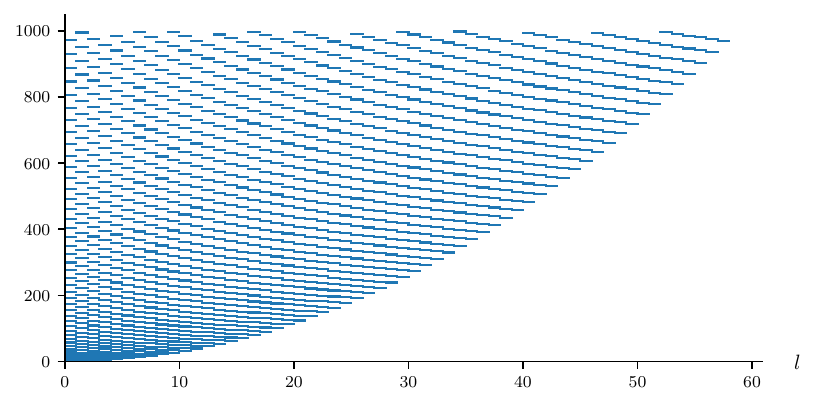}}
\caption{Eigenvalues ordered by angular momentum for $\mu <20$ (left) and $\mu < \bar{\mu}=1000$ (right) in \mbox{$d=2$}.
The lowest eigenvalues (in any dimension) are $1/3$ and $1$, corresponding to the zero modes $f(r)$ and $f'(r)$ respectively.
}
\label{fig:S1}
\end{figure}

To evaluate the fluctuation determinant, we follow Br\'ezin and Parisi and compute the generalized eigenvalues $\mu_n$ (not to be confused with the $\lambda_n$ used in the main text) of the operator
\begin{equation}
\bigl[- \nabla^2 + 1 - 3 \mu_n f^2({\bf x})\bigr] \xi_n({\bf x}) = 0. \label{eq:eigenstates}
\end{equation}
In \mbox{$d=2$}, the eigenvalues depend on a radial angular momentum number $l$ and are doubly degenerate for $l\neq 0$. Figure~\ref{fig:S1} shows all eigenvalues up to a cutoff of $\bar{\mu}=1000$, computed using the shooting method.

In order to express $D(z)$ in terms of the generalized eigenvalues $\mu_n$, we introduce the Weierstrass product
\begin{equation}
W(z) = \prod_n \left(1 - \frac{z}{\mu_n}\right)e^{z/\mu_n} , \label{eq:defD2}
\end{equation}
where $z=1,1/3$ for longitudinal and transversal fluctuations respectively, and the index $n$ runs over all multiples of degenerate eigenvalues. The exponential factor in the the Weierstrass product is only present for $d\geq 2$ and removes the divergent linear-in-$z$ term. This procedure agrees with the counterterm renormalization applied in the main text. We then have
\begin{equation}
\begin{aligned}
    \Omega(z) \equiv \lim_{z\to \mu_0} \frac{W(z)}{(1-z/\mu_0)^{\text{d}(\mu_0)}} = D(\mu_0) \lim_{z\to \mu_0} \left[ \frac{\det\, \Delta_z}{(1-z/\mu_0)^{\text{d}(\mu_0)}}\right],
\end{aligned}
\end{equation}
where $\text{d}(\mu_0)$ denotes the degeneracy of the zero mode with generalized eigenvalue $\mu_0$, and the term in brackets is to be evaluated in the subspace of zero modes. For $\mu_0=1, 1/3$ the normalized zero modes were stated in the main text. We obtain
\begin{equation}
\begin{aligned}
D(1) =  \Omega(1) \left[\frac{dI_4}{4(I_6 - I_4)}\right]^d & \approx \Omega(1) \cdot 0.06022, \\
D\left(1/3\right) = \Omega\left(1/3\right) \left[1-\frac{d}{4}\right] &= \Omega\left(1/3\right) / 2, 
\end{aligned}
\label{irregularZ}
\end{equation}
where the numerical values are for $d=2$.

\begin{figure}[t!]
\includegraphics{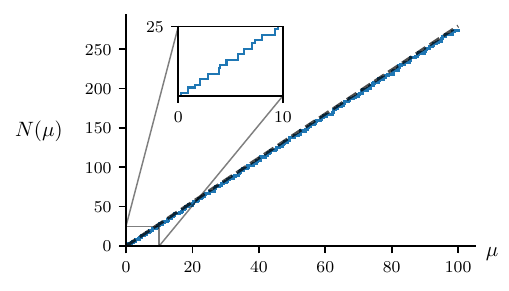}
\caption{
Integrated density of states $N(\mu)$ in $d=2$. The blue curve is the exact results obtained from Eq.~\eqref{eq:eigenstates} and the black dotted line is the asymptotic Thomas-Fermi limit~\eqref{eq:thomasfermi}. The inset highlights the small-$\mu$ behaviour.
}
\label{fig:S2}
\end{figure}

In practice, we evaluate Eq.~\eqref{eq:defD2} by explicitly computing a subset of eigenvalues, $\mu_n \leq \bar{\mu}$, and estimating the contribution of the remaining large-$\mu$ terms using an asymptotic approximation for the spectrum of eigenvalues:
\begin{equation}
\begin{split}
W(z) &= \exp\left[\sum_{\mu_n \leq \bar{\mu}} \left\{\ln \left(1 - \frac{z}{\mu_n}\right)+ \frac{z}{\mu_n}\right\}  + \int_{\bar \mu}^\infty d\mu \, \rho(\mu) \left\{\ln\left(1-\frac{z}{\mu}\right) + \frac{z}{\mu}\right\}\right], \label{eq:defDR}
\end{split}
\end{equation}
where the summation over eigenvalues is replaced by an integral over the asymptotic eigenvalue density~$\rho(\mu)$. This asymptotic density of states $\rho(\mu)$ follows from a Thomas-Fermi approximation: 
The number of states for a deep potential $V({\bf x}) = - 3 \bar{\mu} f^2({\bf x})$ is given by
\begin{equation}
\begin{split}
N(\bar{\mu}) \underset{\bar{\mu}\to \infty}{\sim} &\,  \frac{-(4\pi)^{-d/2}}{\Gamma(d/2+1)}\int d^d x \, V(\bm{r}) 
= \dfrac{(3/4\pi)^{d/2}I_d}{\Gamma(d/2 + 1)} \bar{\mu}^{d/2}  = C_d \bar{\mu}^{d/2} ,
\end{split}
\label{eq:thomasfermi}
\end{equation}
where numerically $C_2 \approx 2.793$. Figure~\ref{fig:S2} compares the exact eigenvalue distribution as determined from solving Eq.~\eqref{eq:eigenstates} with the asymptotic Thomas-Fermi result~\eqref{eq:thomasfermi}. 

With the change of variables $\rho(\mu) d\mu  = dN$, Eq.~\eqref{eq:defDR} becomes
\begin{equation}
W(z) \underset{N\to \infty}{\sim} \left[ \prod_{n=1}^N \left(1 - \frac{z}{\mu_n}\right) e^{\frac{z}{\mu_n}} \right] e^{- \frac{1}{2} \frac{C^2}{N} z^2} .
\label{eq:fullbetter}
\end{equation}
The resulting values for $d=2$ are $D(1)\approx-5.35$ and $D(1/3)\approx1.0558$, in agreement with the ones obtained using the Gel'fand-Yaglom formalism (c.f., Tab.~\ref{main-tab:1} in the main text).

\section{Numerical evaluation of the functional determinant}\label{appendices:R-nitty-gritty}

In this supplemental section, we list useful numerical considerations for the evaluation of Eq.~\eqref{main-eq:det_final} in the main text.

\smallskip

\textit{Initial conditions.} The point \mbox{$r=0$} is singular in Eq.~\eqref{main-eq:R_ODE}, so the numerical integration starts at a small positive value of $r$. To improve the accuracy, we compute the first correction to the initial conditions. By expanding Eq.~\eqref{main-eq:R_ODE} in powers of $r$, and solving the resulting system up to second order, we obtain
\begin{equation}
\mathcal{R}^{(l)}_{z, \varepsilon}(r) = 1 -\frac{3 f(0)^2 z}{2 (d+2 l)} r^2 + \mathcal{O}(r^3). \label{eq:good_boundary}
\end{equation}

\begin{figure}[t!]
\hspace{-20pt}\includegraphics{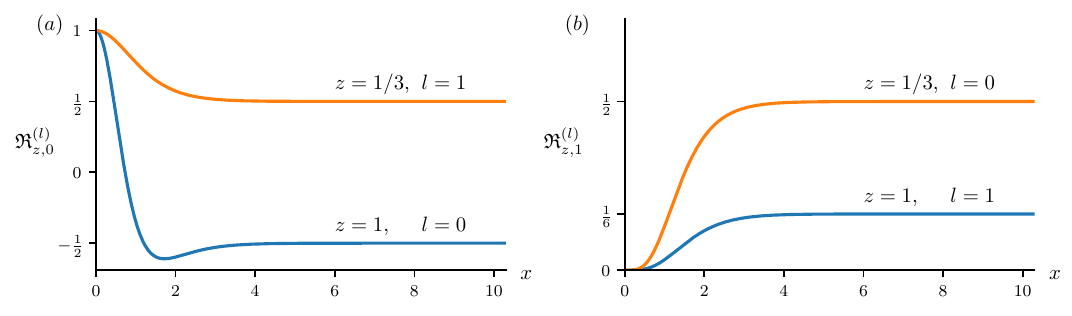}
\caption{Solutions of Eq.~\eqref{eq:zero_R} for $d=1$. In one dimension, $l$ takes only two values, corresponding to even ($l=0$) and odd ($l=1$) eigenfunctions. Panel (a) corresponds to values of $z,l$ that do not contain a zero mode, and their contribution to $D(z)$ is given by $\mathfrak{R}_0$; the values plotted in panel (b) contain zero modes, and their contribution is given by $\mathfrak{R}_1$.
}
\label{fig:S3}
\end{figure}

\textit{Zero modes.} The limit $\varepsilon\to 0$ required to remove the zero mode contribution for \mbox{$l=0$} and \mbox{$l=1$}, is not suitable for numerical implementation. Noting that it can be written in the equivalent form:
\begin{equation}
\lim_{\varepsilon\to 0} \frac{1}{\varepsilon}\mathcal{R}^{(l)}_{z, \varepsilon}(\infty) =\frac{\partial}{\partial \varepsilon} \mathcal{R}^{(l)}_{z, \varepsilon}(\infty) \biggr\rvert_{\varepsilon=0} \equiv \mathfrak{R}^{(l)}_{z, 1}(\infty), \label{eq:R_k2}
\end{equation}
we look for an expression for $\mathfrak{R}_1$ instead (the subindex of $\mathfrak{R}$ indicates the order of the expansion in powers of $\varepsilon$). Substituting $\mathcal{R}^{(l)}_{z, \varepsilon}(r) = \mathfrak{R}^{(l)}_{z, 0}(r) + \varepsilon \mathfrak{R}^{(l)}_{z, 1}(r) + \mathcal{O}(\varepsilon^2)$ into Eq.~\eqref{main-eq:R_ODE} and grouping the corresponding powers of $\varepsilon$, we~obtain
\begin{align}
    \mathfrak{R}_0 ''(r)  + \left[ \frac{2l + d  - 1}{r} + 2 \frac{I_{l+\frac{d}{2}}\left(r\right)}{I_{l+\frac{d}{2}-1}\left(r\right)} \right] \mathfrak{R}_0 '(r) +  3zf^2(r) \mathfrak{R}_0 (r) & = 0 \label{eq:zero_R} \\
    \mathfrak{R}_1 ''(r)  + \left[ \frac{2l + d  - 1}{r} + 2 \frac{I_{l+\frac{d}{2}}\left(r\right)}{I_{l+\frac{d}{2}-1}\left(r\right)} \right] \mathfrak{R}_1 '(r) +  3zf^2(r) \mathfrak{R}_1 (r) & = \frac{r I^2_{\frac{d}{2}+l}(r)-I_{\frac{d}{2}+l-1}(r) \left[2 I_{\frac{d}{2}+l}(r)+r I_{\frac{d}{2}+l+1}(r)\right]}{I^2_{\frac{d}{2}+l-1}(r)}\mathfrak{R}_0(r). \nonumber
\end{align}
Since the boundary conditions \eqref{eq:good_boundary} are independent of $\varepsilon$, we have for $r\ll 1$
\begin{equation}
    \mathfrak{R}_0(r) = 1 -\frac{3 f(0)^2 z}{2 (d+2 l)} r^2, \quad \mathfrak{R}_0'(r) = -\frac{3 f(0)^2 z}{(d+2 l)} r, \quad \mathfrak{R}_1(r) = 0, \quad \mathfrak{R}_1'(r) = 0.
\end{equation}
The solution of Eq.~\eqref{eq:zero_R} for the one-dimensional case is sketched in Fig.~\ref{fig:S3}.

\textit{Angular momentum cutoff.} For $d\geq 2$, the sum~\eqref{main-eq:det_final} over the angular index $l$ needs to be extrapolated to $l\to\infty$. To this end, we consider the asymptotic expansion of the summand $\ln \mathcal{R}^{(z)}_{l,0}(\infty)$ for large values of  $\nu = l + \frac{d}{2} - 1$~\cite{Dunne2006} 
\begin{equation}
\log \mathcal{R}^{(z)}_{l, 0}(\infty) = \frac{V_1}{2\nu}  - \frac{V_3}{(2\nu)^3} + \mathcal{O}(\nu^{-5}), 
\end{equation}
where 
\begin{equation}
\begin{split}
V_1 &= -3z\int_0^\infty dr \, r f^2(r), \\
V_3 &= -3z\int_0^\infty dr\, r^3 f^2(r)\left[2 - 3z f^2(r)\right].
\end{split}
\end{equation}
Using Eq.~\eqref{main-eq:counterterm_def}, the leading large-$l$ behavior for the general term of the sum \eqref{main-eq:det_final} is then
\begin{equation}
\log \mathcal{R}^{(z)}_{l, 0}(\infty) - z r_l \sim - \frac{1}{(2\nu)^3}\int_0^\infty dr\, r^3 (3zf^2)^2.
\end{equation}
The resulting sum over $l$ (including the degeneracy) is evaluated in terms of the polygamma function.

\bibliography{bib_disorder}